\documentstyle[amsbsy,amstex,aps,epsf,multicol]{revtex}

\def\eqref#1{Eq.~(\ref{#1})}

\def\phi{\varphi}

\def\({\left(}
\def\){\right)}
\def\[{\left[}
\def\]{\right]}
\def\<{\left\langle}
\def\>{\right\rangle}
\def\<{\left\langle}
\def\>{\right\rangle}

\def\bea{\begin{eqnarray}}
\def\eea{\end{eqnarray}}

\makeatletter

\def\vlp{\mathopen{\boldsymbol{(}}}    
\def\vrp{\mathclose{\boldsymbol{)}}}   

\def\8{\infty}

\def\undertext#1{\vtop{\hbox{#1}\kern 1pt \hrule}}

\def\VEV#1{\left\langle\,#1\,\right\rangle}

\def\be{\begin{equation}}
\def\ee{\end{equation}}
\def\bea{\begin{eqnarray} & &}
\def\eea{\end{eqnarray}}

\def\rf#1{(\ref{#1})}

\def\t{\tilde}

\title{Integer Quantum Hall Transition and Random $SU(N)$ Rotation}
\author{Stanislav~Boldyrev and Victor~Gurarie
\\
{\em 
Institute 
for Theoretical Physics, 
Santa Barbara, California 93106}}

\date{\today}

\begin{document}

\input psfig.sty



\maketitle

\begin{abstract}
We reduce the problem of integer quantum Hall transition to  
a random rotation of an~$N$-dimensional vector by
an~$su(N)$ algebra, where only~$N$ specially selected 
generators of the algebra are nonzero. The group-theoretical
structure revealed in this way allows us to obtain a new series
of conservation laws for the equation describing the electron density
evolution
in the lowest Landau level. The resulting formalism is particularly
well suited to numerical simulations, allowing us to obtain the
critical exponent $\nu$ numerically in a very simple way. 
We also suggest that if the number of 
nonzero generators 
is much less than~$N$,
the same model, in a certain intermediate time interval,  describes  
percolating properties of a random incompressible
steady two-dimensional 
flow. In other words, quantum Hall transition in a very smooth
random potential inherits certain properties of percolation. 

\vspace{5mm} \noindent
\end{abstract}
\vspace{1mm}
\begin{multicols}{2}


{\bf 1}. The quantum Hall transition is a delocalization transition of a 
particle moving on a two dimensional plane with a random potential and a
strong magnetic field perpendicular to the plane \cite{review}. 
Recently, an elegant approach to this
transition was suggested by Sinova,
Meden, and Girvin~\cite{Girvin}. The idea of the method is to
consider the quantum mechanical states of a particle belonging to
the lowest Landau level, and to project the density 
operators~${\hat \rho}(x,t)$ onto these functions. The
spatial correlation properties of the particle can then be described 
by the correlation function of these projected density
matrices,~$G(x,t)=\mbox{Tr}\vlp {\hat \rho}(x,t) 
{\hat \rho}(0,0)\vrp $ or by
its average, $\langle G(x,t) \rangle$.
 In this formula, ``Tr" corresponds to summing over all 
the states of the lowest Landau level, and the angular 
brackets denote averaging over the random potential. 

In a certain 
sense, $G(x,t)$~is the probability of 
transition of a quantum particle from the origin to point~$x$ in 
the course of time~$t$. This statement would be exactly true if we
allowed the particle to travel over all the quantum states, i.e., not 
only those belonging to the lowest Landau level. Not having the full 
system of wave functions does not allow us to localize the particle 
at a
distance smaller than the magnetic length~$l^2=\hbar c/(e B)$. 
For example,  the 
initial 
condition for the~$G$ function have the 
form~$G(x,0)=A\exp(-x^2/2l^2)$,  with some normalization constant~$A$. 
However, for strong enough magnetic 
fields, the magnetic length~$l$ is mush smaller than the size of 
the system and particles can be adequately localized.

The projected
density operators obey the  Schr\"odinger equation~
\begin{eqnarray}
i{\hbar
}\frac{\partial}{\partial t}
{\hat \rho}=[{\hat H},{\hat \rho}],
\label{schroedinger}
\end{eqnarray} 
whose Hamiltonian is
just the random potential~$V(x)$ projected onto the lowest Landau
level,~${\hat H}=\int \mbox{d}^2\,x~V(x){\hat \rho}(x)$. 
The correlation function~$G$ satisfies an analogous equation,
which due to remarkably
simple commutation relations between the density operators $\hat \rho$, 
can be written in the Fourier space ($k$ is a two-dimensional
vector, $k=(k_1, k_2)$) \cite{Girvin}:
\begin{eqnarray}
i\hbar \frac{\partial}{\partial t} G(k,t)&=
&\int\mbox{d}^2q \,2i\,\sin\left(
\frac{l^2}{2}k\times q\right) V(k-q) \times \nonumber \\
&{}&\exp\left[
-\frac{l^2}{2}\left(k^2-k\cdot q \right)\right]G(q,t).
\label{g_function}
\end{eqnarray}
For a simple,
``from the first principles," 
derivation of this equation we refer the reader 
to~\cite{Gurarie_Zee}. All the information about the quantum Hall transition
(QHT) 
is 
contained in this equation, which we use as a starting point 
in the present work. Unfortunately, it is not obvious how to 
solve this equation.

In this letter, we are going to demonstrate the 
hidden group-theoretical structure of Eq.~(\ref{g_function}). We 
show that it admits a series of conserved integrals of motion that
have not been known before. We reduce the problem of QHT
to a finite-dimensional problem of a random rotation by
banded $su(N)$~matrices with the band width $\sim \sqrt{N}$.
We also show that when the width of the band is
much smaller than~$\sqrt{N}$, such a model describes classical
percolation.

Let us start with the classical limit~$l \rightarrow 0$. In this case,
 Eq.~(\ref{g_function}) is simplified considerably,
\begin{equation}
\label{perc}
{\partial \over \partial t} G(x,t) = \epsilon_{ij} \partial_i V(x)
\partial_j G(x,t). 
\end{equation}
A formal solution to~(\ref{perc}) can be written at
once~\cite{Gurarie_Zee}. It is $G(x,t) = \delta(x-x(t))$ where
${dx_i \over dt} = \epsilon_{ij} \partial_j V(x)$, the equations which
were thoroughly studied in \cite{Fevers}. 
From here it is easy to deduce that we are trying to describe
a particle which percolates along the equipotential lines of a random
potential. Obviously that does not capture the physics of quantum Hall
transition.

The reason why a solution to~(\ref{perc}) was so easy to write down
is contained in its infinitely many integrals of motion. Indeed, it is 
not hard to check that any expression of the form
\begin{equation}
\label{integrals}
I_{mn}=\int d^2 x \ G^m V^n
\end{equation} is conserved along the solutions of~(\ref{perc}).
The conservation of~$I_{10}$ and~$I_{11}$ is simply a consequence
of the conservation of probability and energy. However, 
higher order integrals
have a less obvious meaning. 
Now, in a quite remarkable way, these integrals of motion do not get
destroyed as one goes back to the original equations (\ref{g_function}),
at least for integer~$m$ and~$n$. They only get slightly modified.

To see that, let us put the particle on a torus. The
wave functions of the lowest Landau level can be written explicitly
in the Landau gauge,
\begin{eqnarray}
\label{wavef}
\psi_\alpha(x,y)&=&\left[ \sum_m \exp  \biggl( 2 \pi (x+i y) (N m + \alpha) 
\biggr.
\right. \nonumber \\
 &-&\Biggl. \left.  {(N m +\alpha)^2
\over N} \pi \right) \Biggr] e^{-\pi N x^2}.
\end{eqnarray}
Here~$0\le x,y \le 1$, $N$ denotes the number of
flux quanta through the torus and~$\alpha$ goes from~$0$ to~$N-1$, labeling
the~$N$ states on the torus in the lowest Landau level. These
wave functions describe the electron localized along a narrow
strip around the line $x=\alpha/N$. Notice that the magnetic length~$l$ 
is now
automatically chosen in the form 
\begin{equation}
\label{magneticlength}
l^2=\frac{1}{2 \pi N}. 
\end{equation}

It is now a matter of simple calculations to project
the density operators $\hat \rho_{(k_1,k_2)} \equiv \exp 
\left( 2 \pi i \left(k_1 x+k_2 y \right)\right)$ 
onto the lowest Landau level
on the torus. In these notations, $k_1$ and $k_2$ are integer 
numbers. As a result, the density operators
are now $N \times N$ matrices in the basis of states
\rf{wavef} which 
can be written in the following form. 

Consider a pair of unitary
unimodular $N\times N$~matrices:
\begin{eqnarray}
\label{matrices}
h=\left( \begin{array}{ccccc}
0     &   1  &\dots &\dots & 0    \\
0     &   0  & 1    &\dots & 0    \\
\dots &\dots &\dots &\dots &\dots \\
1     &\dots &\dots &\dots & 0
\end{array} \right),
\end{eqnarray}
\begin{eqnarray}
f=\mbox{diag}(1,\epsilon,\dots,\epsilon^{N-1}),
\end{eqnarray}
where~$h$ is a cyclic permutation matrix, and~$\epsilon=\exp(2\pi
i/N)$. These matrices have the following properties:
$\label{matrix_properties}
hf=\epsilon fh,\quad h^N=f^N=1$.
Now introduce the matrices 
\begin{equation} 
L_{(k_1,k_2)} = \epsilon^{k_1 k_2/2} f^{k_1} h^{k_2},
\end{equation}
where~$(k_1,k_2)\neq(0,0)$.
The matrices~$L_{(k_1,k_2)}$ are periodic in~$k_1$ and~$k_2$ with
period~$N$ up to  coefficient~$\pm1$. They can be chosen as a 
basis for the~$su(N)$ algebra.
The product and the commutator of two such operators have the
following form,
\begin{eqnarray}
\label{l_multiplication}
L_q L_p & = & \exp \left( \frac{\pi i}{N} q 
\times p\right) L_{q+p}, \\
\left[ L_q , L_p \right]  &=& 
2i\sin \left( \frac{\pi i}{N} q\times p
\right) L_{q+p}.
\label{l_commutation}
\end{eqnarray}
One can easily
write down the explicit expression for the matrix elements of these
matrices:
\begin{eqnarray}
\label{explicit_generators}
\left(L_{(k_1,k_2)}\right)_{\alpha,\beta}=\epsilon^{k_1k_2/2
+k_1(\alpha-1)}\delta_{\alpha, \beta -k_2}\vert_{\mbox{mod}\, N}.
\end{eqnarray}
Then the density operator can be expressed in terms of these 
matrices as
\begin{equation}
\label{density}
\hat \rho_{(k_1,k_2)} = \exp \left({- {k_1^2 + k_2^2 \over 2 N} \pi} 
\right)
L_{(k_1,k_2)}.
\end{equation}
To be specific, \rf{density} gives the density operator in the 
Schr\"odinger representation, as opposed to \rf{schroedinger} where
the Heisenberg representation was assumed. From this point on,
we will understand $\hat \rho$ as a time-independent density operator. 

It is not difficult to check that the density operator written in 
this form
indeed satisfies the commutation relations discussed, for example,
in~\cite{Girvin} with the magnetic length chosen according
to~\rf{magneticlength}.

The projected Hamiltonian takes the form
\begin{equation}
\label{ham}
\hat H=\sum_{k_1 k_2} \ V(-k_1, -k_2) \hat \rho_{(k_1,k_2)}.
\end{equation}
Since the Fourier components of the potential~$V(x)$ are random, 
we see that the Hamiltonian is none other than a random~$su(N)$ 
matrix. 
This completes the projection to the lowest Landau level on the torus. 
At this point we can easily rederive~(\ref{g_function}) simply by
writing~$\hat \rho$ as a matrix and commuting it with the
Hamiltonian~$\hat H$ as prescribed in~(\ref{schroedinger}) 
using the explicit matrix definitions~\rf{density}. 
Finally, let us introduce the matrix $\hat G=\sum_{k_1 k_2} G(k,t)
\hat \rho_{(k_1,k_2)}$. Here as a consequence of choosing $k_1$ and
$k_2$ on the torus as integer numbers, we denote $k=2 \pi (k_1,k_2)$.
As a consequence of~(\ref{g_function}), it
satisfies the equation identical to~(\ref{schroedinger}),
\begin{eqnarray}
i{\hbar
}\frac{\partial}{\partial t}
{\hat G}=[{\hat H},{\hat G}].
\label{schroedinger1}
\end{eqnarray} 

Through all these manipulations we succeeded in reducing the original
problem of a particle on an infinite plane in a random potential
to a problem with a finite number of degrees of freedom. That
allows us immediately to write down the generalizations of the
integrals of motion~(\ref{integrals}). Indeed, since a trace
of a commutator of finite matrices is equal to zero,
\begin{equation}
\label{quantumintegrals}
\t I_{mn} = {\rm Tr} \left( \hat G^n \hat H^m \right)
\end{equation} 
is conserved. These integrals can be expressed in terms of
$G(k,t)$ and $V(k)$ using explicit expressions for the
matrices $\hat \rho$. 

The first few of the integrals in~\rf{quantumintegrals} 
coincide with their ``classical'' counterparts in~\rf{integrals}. 
For example, $\t I_{10}$ and~$\t I_{11}$ are equal 
to~$I_{10}$ and~$I_{11}$ and are still the probability 
and energy conservation, respectively. Higher order integrals of
motion become increasingly more complicated. For example, 
\begin{eqnarray}
I_{30} & = & 
\sum_{k,s,r}\,g(k)\,g(s)\,g(r-k-s)\ 
\nonumber \\
&\times & \exp\left[\frac{\pi i}{N}(k\times
s+k\times r+ s\times r)+ \pi i N r_1 r_2\right],
\label{explicit_integral}
\end{eqnarray}
where $k=2 \pi (k_1,k_2)$, $s=2 \pi (s_1,s_2)$, $r=2 \pi N (r_1, r_2)$,
$g(k)=G(k)\exp[-\pi \left( k_1^2+k_2^2\right)/(2N)]$, 
and the summation is going over integer $k_1$, $k_2$, $s_1$, $s_2$, $r_1$, and~$r_2$.

Despite being rather complicated, this expression reduces
to its classical counterpart, $I_{30} = \int d^2 x \ G^3(x)$,
in the limit~$l \rightarrow 0$ ($N \rightarrow \infty$). 
It is of course possible to show, 
after
some algebra, that it is indeed conserved under the time
evolution~\rf{g_function}. A simple mathematical reason lies behind this: 
in the limit~$N\to
\infty$, the operators~$\hat \rho$ become the generators of the group
of volume preserving diffeomorphisms on a
torus~\cite{Arnold_Khesin,Zeitlin1}. Such a group
represents motion of an incompressible fluid, which is
precisely the meaning of~(\ref{perc}).
\vskip3mm

{\bf 2}. 
We are going to demonstrate now that 
all the features of QHT are preserved in this
picture. Consider a particle placed in one of the states~\rf{wavef}.
That means, it is localized along the $x$~direction and is extended
along the $y$~direction. Wait some time~$t$ and measure the average
square of the 
displacement of the particle along the $x$~direction,
\begin{equation}
\label{x2}
\VEV{x^2(t)}\propto {1 \over N^3} 
\sum_{\alpha,\beta=1}^{N} \left| \left(e^{{i \over \hbar } \hat
H t } \right)_{\alpha \beta} \right|^2(\alpha-\beta)^2,
\end{equation}
where~$()_{\alpha \beta}$ denotes the matrix element of the matrix inside
the brackets. 
We expect that, according to~\cite{Girvin,Gurarie_Zee}, at large
enough times but before the finite size of the system is reached,
the deviation~$\VEV{x^2(t)}$ behaves as 
\begin{eqnarray}
\langle x^2\rangle \sim  t^{1-1/(2\nu)}. 
\label{deviation}
\end{eqnarray}
The critical exponent~$\nu$ is to be determined, 
and is believed to be close to~$7/3$.

The Hamiltonian~\rf{ham} appears to be a linear combination of~$su(N)$ 
matrices with random coefficients. However, if it were indeed a 
random~$su(N)$ matrix with probability distribution 
{\em invariant} under~$SU(N)$ rotations, it would lead to a
particle instantaneously hopping all over the torus. In fact, such a
Hamiltonian would be unphysical. As a consequence of the strong
exponential suppression of high Fourier modes in the density
operators~\rf{density}, the~$SU(N)$ rotation invariant Hamiltonian 
corresponds
to the random potential~$V(x)$ varying considerably at distances
much smaller than the magnetic length. In fact, in the continuum limit
it would lead to the potential having Fourier harmonics growing 
exponentially at large~$k$, as follows from the analysis
of~\rf{density}.

A more physical setting involves the potential whose Fourier
harmonics at least do not grow at large~$k$. While it is possible
to choose a particular random~$V(x)$ satisfying this property and 
compute
the Hamiltonian~\rf{ham}, it is not necessary to do so. 
Instead, we can just point out that the exponential in~$\rf{density}$
strongly suppresses high Fourier harmonics of the random potential,
such that~$k_1, k_2 >n= \sqrt{2N/\pi}$.  
Therefore, we can simply choose the Hamiltonian to be
\begin{equation}
\label{hamnew}
\hat H = \sum_{k_1,k_2=0}^{n} 
v_{(k_1,k_2)} L_{(k_1,k_2)},
\end{equation}
where $v_{(k_1,k_2)}$ are random independent variables with 
equal mean square values $\VEV{v^2}$. 
The crucial part of~\rf{hamnew} is that the Hamiltonian is
a linear combination of only~$N$ generators of the~$su(N)$ algebra,
out of total~$N^2$ generators, with random coefficients. From
the explicit form~\rf{explicit_generators}
of matrices~$L_{(k_1,k_2)}$ we see that the
random matrix~$\hat H$
has nonzero elements only on the diagonal strip of the 
width~$\sim N^{1/2}$. 
Within the strip, only the matrix elements no farther 
than~$\sim N^{1/2}$
from each other along a given diagonal are correlated with each other. 
Such {\it banded random matrix} theory, which reproduces the
physics of the QHT, was considered in a similar context in 
\cite{Mieck}, although
the Hamiltonian \rf{hamnew} was not introduced there.

This framework provides a very convenient setting for 
numerical calculation
of~$\nu$. 
Below we 
present the numerical simulations for matrix 
1000$\times$1000. 
The Hamiltonian has been chosen in the form~(\ref{hamnew}).
The random 
coefficients~$v_{(k_1,k_2)}$
have thus been generated in the 
square~$|k_1|,|k_2|<n=\sqrt{2N/\pi}$, and the
real and imaginary parts of each Fourier 
mode~$v_{(k_1,k_2)}=v^*_{(-k_1,-k_2)}$ were chosen
randomly, independently, and uniformly from 
the interval~$[-0.5,\,0.5]$.
Instead of calculating the matrix exponent in~(\ref{x2}) directly,
we have simulated the equation
\begin{eqnarray}
\dot Z_{\alpha}=i {\hat H}_{\alpha \beta} Z_{\beta},
\label{z_function}
\end{eqnarray}
where~$Z_\alpha$ is a wave function in the representation of 
the states on
the torus. It is easy to check that in this representation,
\begin{eqnarray}
g(k_1, k_2)=Z_{\alpha}^*\left[L_{(k_1, k_2)} \right]_{\alpha \beta} Z_{\beta},
\end{eqnarray}
where~$g(k)$ is introduced in~(\ref{explicit_integral}).

After the random potential~$\hat H$ had been 
generated, and the initial distribution had been chosen in the form
\begin{eqnarray}
Z_{\alpha}=\left\{
\begin{array}{lc}
1, & \mbox{if} \quad \alpha=\alpha_0, \\
0, & \mbox{otherwise},  \\
\end{array}\right.
\end{eqnarray}
the dispersion~$\VEV{x^2(t)}=
\sum_{\alpha} (\alpha-\alpha_0)^2 |Z_{\alpha}|^2$ was
calculated as a function of time. The same calculation was 
performed for all the initial positions of the particle, 
$\alpha_0=1,\dots,N$, and the average was taken over all 
such realizations. The result is shown 
in~Fig.~(\ref{dispersion}). As we see, we closely 
reproduce the universally accepted value
of~$\nu$ without much difficulty. 
{
\columnwidth=3.4in
\begin{figure} [tbp]
\centerline{\psfig{file=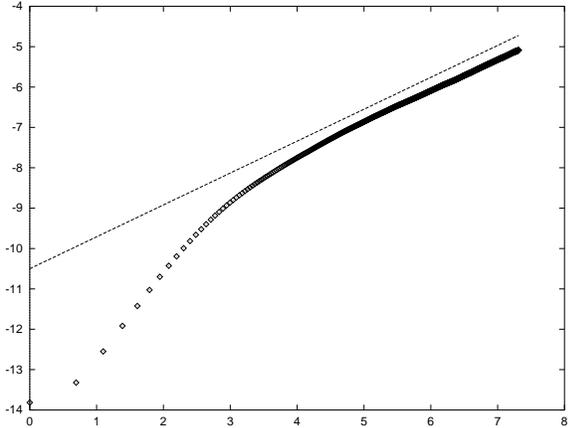,width=3.2in,angle=-90}}
\vskip3mm
\caption{$\mbox{Log}(\langle x^2 \rangle )$ is plotted 
as a function 
of~$\log(t)$ for
$N$=1000,~$n=15$. The light line has the slope~$0.79$ which
corresponds to~$\nu=2.38$. }
\label{dispersion}
\end{figure}
}
\vskip3mm
{\bf 3}. As was already discussed, a naive~$l \rightarrow 0$ limit 
leads
to the percolation picture of QHT. However,
$l \rightarrow 0$ is the same as~$N \rightarrow \infty$. It should
indeed be possible to reproduce the percolation behavior if the scale
of the random potential is much larger than the magnetic 
length~$l$, but much smaller than the size of the system. In 
other words, we need to keep only such modes in~\rf{ham} for 
which~$1\ll n \ll N^{1/2}$. The percolating behavior holds until the
width of the diagonal strip of the matrix~$\exp(i{\hat H}t)$ becomes
equal to~$N^{1/2}$. After that, the regime changes to QHT. 
The square of this width is given by~(\ref{x2}),
and therefore grows as~$t^{1-1/(2\nu^*)}$, with the percolation~$\nu^*$
equal to~$4/3$. The crossover time (mixing time) is thus equal,
in physical units, 
to~$V_0 t_m/ \hbar\sim
(N/n^2)^{2\nu^*/(2\nu^*-1)}\sim (l_0/l)^{4\nu^*/(2\nu^*-1)}$, 
where~$l_0$ is
the scale of the random potential and $V_0$ is its typical amplitude. 
The percolation regime is valid 
for~$t < t_m$.

Our simulations of
Eq.~(\ref{z_function}) show that for~$n \leq (2N/\pi)^{1/2}$, 
the behavior of~$x^2$
coincides with~(\ref{deviation}) at~$t\to \infty$, but follows 
some intermediate asymptotics before that. We expect  
these asymptotics to be identical to those describing
steady percolating flow of an incompressible two-dimensional fluid, 
i.e.,~$\nu^* = 4/3$. This seems to agree with our numerics, 
but more extensive simulations are required~\cite{Boldyrev_Gurarie}. 
In the other 
limit,~$n \gg \sqrt{2N/\pi}$, the behavior  
becomes diffusive, $\langle \alpha^2 \rangle\sim t$.

In real experiments, the random potential length is often
much larger than the magnetic length $l$. It would be interesting
to devise an experiment which would probe the intermediate percolation
asymptotics, perhaps by looking at finite frequency conductivity.

\vskip3mm
In conclusion, we have presented a model that 
reduces the problem of quantum Hall
transition to a finite-dimensional problem of random rotation 
by~$su(N)$ matrices. As a consequence, the 
equation~\rf{g_function} describing quantum Hall transition, 
admits a series
of integrals of motion. The random rotation matrix~$\hat H$ 
is quite arbitrary, except for the dependence 
on the parameter~$N$ through the exponential cutoff 
in~\rf{density}. This rather crucial 
dependence amounts to
vanishing of all the matrix elements but those
belonging to the diagonal strip of the width~$n\sim N^{1/2}$. 
Within this framework, the quantum Hall transition exponent~$\nu$
can easily be evaluated numerically. 
Attempts to change~$n$ independently of~$N$ 
lead to other models, e.g. those describing diffusion and 
percolation in a steady incompressible 2D~flow. That leads to
a prediction that in very smooth potentials quantum Hall transition
should exhibit some properties of percolation. 

\vskip3mm

We are grateful to A.~Zee, J.~Chalker, I.~Gruzberg,   
A.~W.~W.~Ludwig, and A. Kamenev for important discussions.
This work was supported by the NSF grant PHY 94-07194

\end{multicols}
\end {document}